\documentclass[12pt,preprint]{aastex}

\def\kms  {km~s$^{-1}$}
\def\masy {mas~y$^{-1}$}
\def\uasy {$\mu$as~y$^{-1}$}
\def\uas  {$\mu$as}
\def\uJy  {$\mu$Jy}
\def\etal {et al.~}
\def\eg   {e.g.~}
\def\ie   {i.e.~}
\def\hho  {H$_2$O}

\def\Vlsr {\ifmmode {V_{\rm LSR}}\else {$V_{\rm LSR}$}\fi}
\def\Ro   {\ifmmode {R_0}\else {$R_0$}\fi}
\def\To   {\ifmmode {\Theta_0}\else {$\Theta_0$}\fi}

\slugcomment{.}
\shorttitle{Motions of Galaxies in the Local Group and Beyond} 
\shortauthors{Reid \etal}

\begin{document}

\title{Motions of Galaxies in the Local Group and Beyond: \\
       An Astro2010 Science White Paper}

\author{M. J. Reid (Harvard-Smithsonian CfA)}
\author{A. Brunthaler (MPIfR), K. M. Menten (MPIfR), L. Loinard (UNAM), J. Wrobel (NRAO)}  
\author{} 
\author{}
\author{}
\author{}
\author{\epsscale{0.8}\plotone{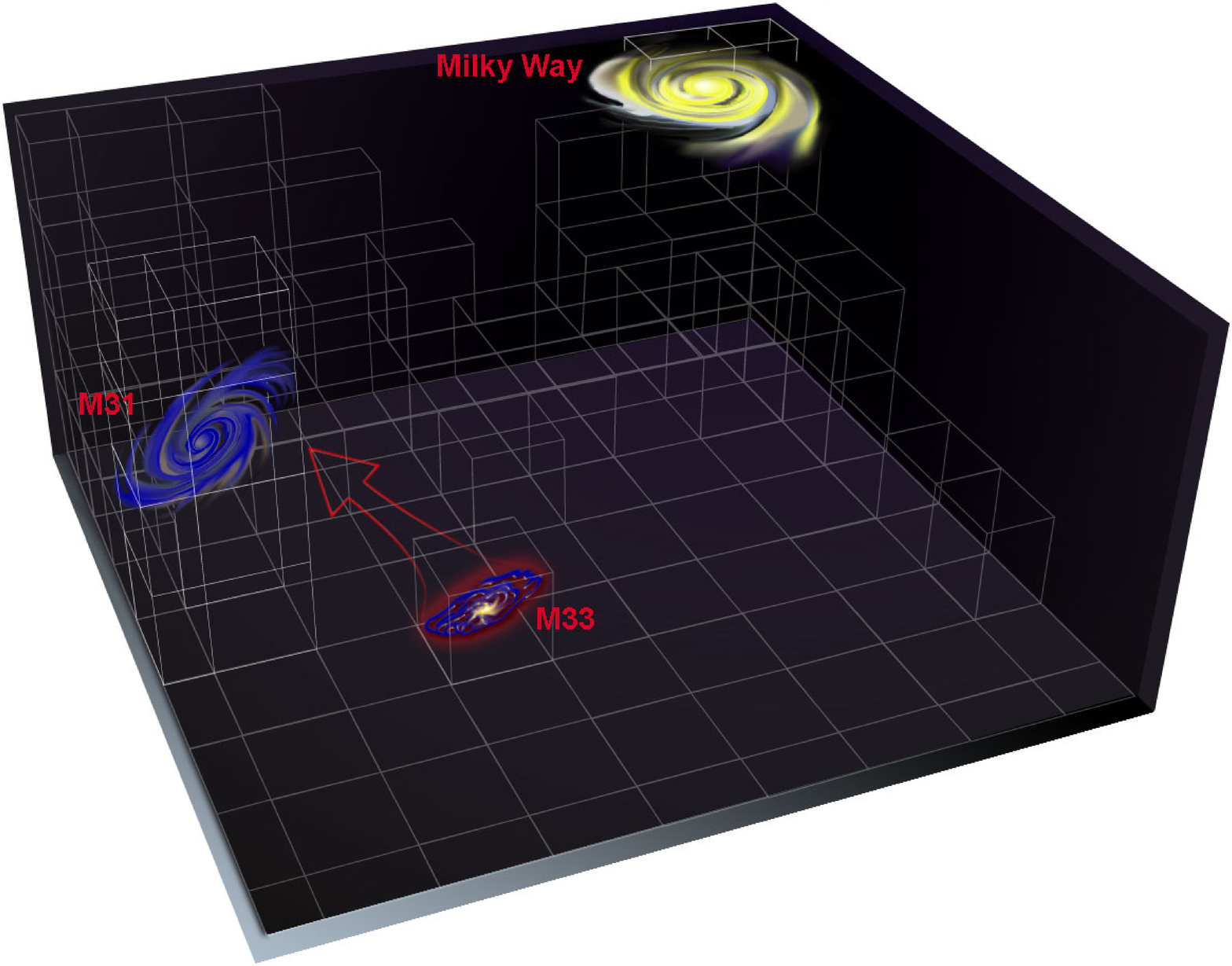}}
\author{}
\author{}
\author{}

\begin{abstract}
Recent advances with the VLBA have resulted in $\sim10$~\uas\ 
astrometry for compact sources in external galaxies, and measurement of the 
proper motion of Local Group galaxies has been demonstrated.  With improved 
telescopes and equipment, we could greatly improve upon and expand these 
measurements, including a measurement of the proper motion of the Andromeda 
galaxy, which is key to understanding the history and fate of the Local Group. 
The combination of optical velocities and radio astrometric data would allow 
detailed modeling of the mass distributions of the disks, bulges, and dark 
matter halos of galaxies in clusters.
\end{abstract}

\section {Background}

The nature of spiral nebulae was actively debated in the 1920's. 
While some astronomers favored a short distance and a Galactic origin, others  
argued for their extragalactic nature.  In 1923, van Maanen measured
photographic plates of M~33 separated by about 12 years and claimed to have 
detected the galaxy rotating at $\sim20$~\masy, clearly requiring a near distance 
\citep{vanMaanen}.  However, a few years later, Hubble discovered Cepheids 
in M~33, providing evidence for a very large distance and requiring it to be 
extragalactic \citep{Hubble}.  At Hubble's distance, the expected proper motions from the 
rotation of M~33 are only $\approx$ 30 \uasy, nearly three orders of magnitude 
smaller than the motions claimed by van Maanen.  (The source of error in van Maanen's 
measurements was never clearly identified.)  After more than 80 years, the goal 
of measuring the rotation and absolute proper motions of galaxies remains 
interesting for our understanding of the dynamics and geometry of the Local Group 
and beyond.

\section {Scientific Context: Extragalactic Proper Motions}

We are currently poised to make truly dramatic progress in understanding 
the dynamics, and hence dark matter mass distribution, of the Local Group.
Over the next 10 to 20 years, we could make measurements with accuracies of
$\sim0.1$~\uasy\ and thus measure the 3-dimensional velocities of galaxies in the 
Local Group and in nearby groups out to the Virgo Cluster.

\subsection{Motions in the Local Group}

The distribution of dark matter in galaxies and groups of galaxies is
one of the major problems in observational cosmology.  The Local Group
provides the nearest and best system for detailed study.  Various attempts
to ``weigh'' the Milky Way and the Andromeda galaxy have resulted in
a large range of values.   The major problem is that most studies
work with only one-dimensional (radial) velocity components, introducing 
significant ambiguities and requiring statistical mass estimators, 
which suffer from small sample sizes and/or unknown biases from 
non-isotropic distributions.  Clearly the most reliable way to derive mass 
distributions is from 3-dimensional velocity measurements through proper motions.   

\begin{figure}[ht]
\epsscale{0.7} 
\plotone{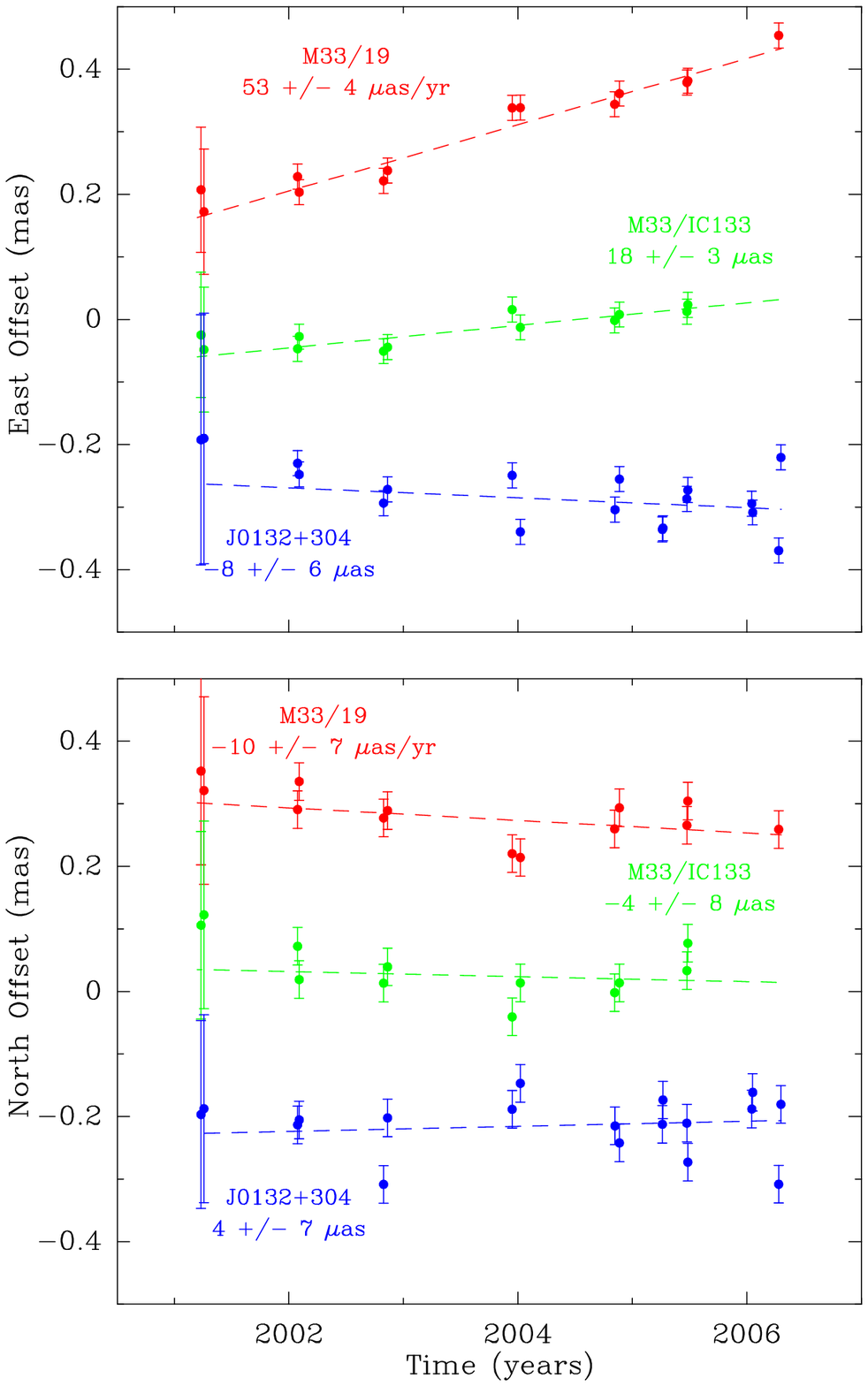}
\caption{
Absolute proper motion data for 2 sources of \hho\ masers on opposite sides of M~33 
relative to a background quasar.  The two panels show motions in R.A. and 
in Dec.  In each panel, data plotted in red ({\it top}) are for the maser M~33/19 
and in green {\it(middle)} are for the maser M~33/IC133.   Also shown in blue 
{\it (bottom)} is a ``control'' measurement of a weak quasar J0132+304 relative to the
background quasar used for the masers.   
Note that even with current VLBI capabilities, proper
motion uncertainties of a few \uasy\ are achieved in only 4 years of observation.
         }
\label{fig:m33}
\end{figure}

Currently, proper motion measurements, done optically, are only possible for close
satellites of the Milky Way (eg, for the LMC see \citet{Kallivayalil:06}). 
The satellites of Andromeda are an oder of magnitude more distant and current optical 
telescopes are inadequate.  Only VLBI has yielded proper motions for the
Andromeda satellites M~33 and IC10 \citep{Brunthaler:05,Brunthaler:07}.  
Fig.~\ref{fig:m33} shows recent astrometric data for two sites of \hho\ masers 
in M~33 relative to a background quasar.  The relative motion of the two masers 
on opposite sides of the galaxy directly yields the angular rotation of the 
galaxy (``solving'' the van Maanen/Hubble debate); when combined with the rotation 
curve from HI mapping, this yields a direct estimate of the galaxy's distance.   
Removing the effects of the galaxy's rotation, the absolute proper motion of 
M~33 is obtained, leading to strong constraints on the dark matter halo and 
motion of Andromeda \citep{Loeb:05}.

\begin{figure}[ht]
\epsscale{0.8} 
\plotone{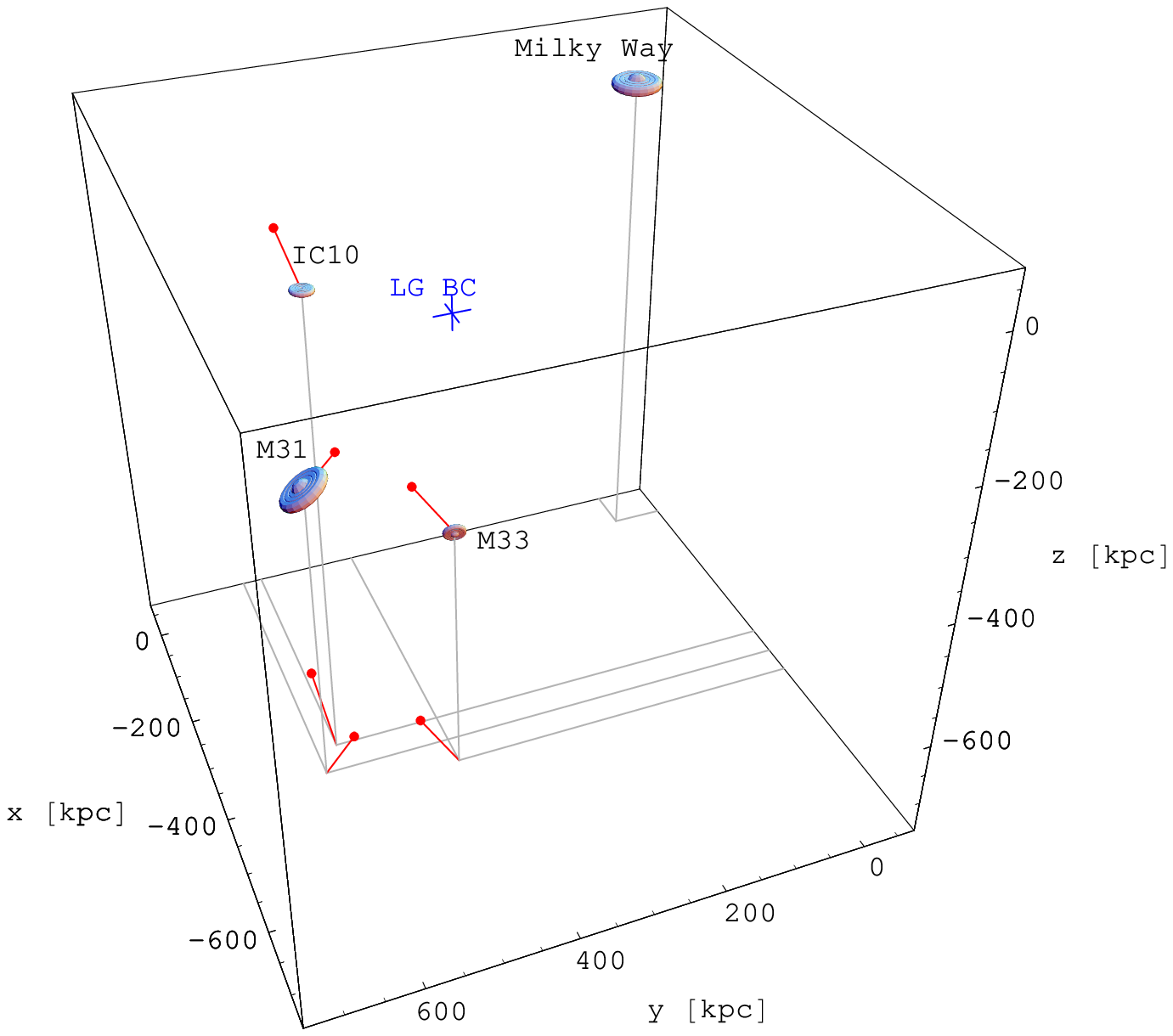}
\caption{{
Local Group proper motions.  Schematic view of the Local Group of Galaxies with the
Milky Way in the upper right and the Andromeda galaxy (M~31) in the left.   The measured
3-D motions of M~33 and IC~10 are indicated with red lines.  Only the known
radial component of the motion of Andromeda is indicated.
Unless the Andromeda galaxy has a 
substantial proper motion ($\sim100$~\kms) or much less dark matter than previously 
estimated \citep{Loeb:05}, M~33 would have ``hit'' Andromeda in the past and would not 
appear as a thin disk today.
}}
\label{fig:local_group}
\end{figure}

The full 3-dimensional motion vectors for two Andromeda satellites, plus the 
1-dimensional motion (radial) of the Andromeda galaxy, are shown in 
Fig.~\ref{fig:local_group}.
The proper motion of the Andromeda galaxy (yet to be directly measured) is key to 
understanding the history and fate of the Local Group, since a potential collision
of the Andromeda galaxy and the Milky Way would ``reconfigure'' the entire 
Local Group.  Depending on the proper motion of Andromeda, the
two dominant galaxies could have a ``head on'' collision, a grazing collision, or no
collision at all.  With a sizeable proper motion of $\sim100$~\kms, depending on the 
total dark matter mass in the Local Group, they could orbit each other or even be unbound. 
A high accuracy measurement of the proper motion of Andromeda is possible in the coming
decade with modest upgrades to the VLBA.

Measurements of the proper motions of more Milky Way and Andromeda satellite galaxies 
(some of which will come from radio interferometry), and improved numerical simulations, 
will be critical to understanding the size and mass of the disk, bulge, and dark matter 
halos of these two dominant galaxies in the Local Group.  
Measuring the proper motions of several maser sources within one galaxy can also give 
a direct measurement of the galaxy's distance (rotational parallax). This method, 
applied to M~31, M~33, and the LMC, could yield distances with accuracies of a few percent.
These {\it direct} geometric distance estimates to galaxies, which are inherently free 
of systematic effects, such as metalicity differences, are of great importance for an
independent re-calibration of extragalactic distance indicators.

\subsection{Motions outside the Local Group}

With astrometry at the \uas\ level one can extend extragalactic proper motion 
studies to galaxies outside the Local Group and address important questions related 
to galaxy evolution and flows.  The M81 group, the nearest group with interacting 
galaxies, is an excellent system to study the evolution of galaxies through interactions. 
In physical contrast, the galaxy flow in the nearby Canes Venatici I cloud appears
to be driven by free Hubble expansion, and that flow can be traced via the proper motions 
of the low-luminosity active galactic nuclei hosted by galaxies in the cloud.  

Another
important target for proper motion studies is the Virgo Cluster, the
nearest and best studied galaxy cluster. While it is much further away
than galaxies in the Local Group, the relative velocities of the
galaxies in the deep gravitational potential in the center of the Virgo
Cluster can be several thousand \kms, an order of magnitude larger than
in the Local Group. Knowledge of the 3-dimensional motions of the galaxies  
can help one to address the mechanisms by which the cluster environment affects 
galaxy properties. For example, it has been well established that cluster spirals
are quite gas poor compared to  spirals in the field.  This is generally
thought to be due to ram-pressure stripping of the galaxy's interstellar medium 
as the galaxy moves through the intracluster medium and is observed
to occur in several Virgo spirals, where large tails of gas are being
torn from the galaxy by interaction with the intracluster medium. 

\section {Telescope Needs}

The observations needed to measure proper motions of Local Group galaxies outlined 
above can be accomplished with the advances outlined in 
Table~\ref{table:telescope_impact}.  Some of the goals can be achieved with 
modest upgrades in data recording equipment at the VLBA.  Increasing the VLBA data 
recording rate by more than two orders of magnitude, from 256 Mbps to 32 Gbps 
(which requires no new technology), would have two dramatic advantages: 
1) increase the signal-to-noise ratio for continuum sources (both AGN
in target galaxies and in background reference quasars) and 
2) improve astrometric accuracy by making far more background quasars available as 
position references.  
Regarding advantage 2), VLBI astrometric observations are often limited by systematics 
that cancel proportionally to the separation of the maser target and background quasar.
The factor of 11 (\ie $\sqrt{32 {\rm Gbps}/256 {\rm Mbps}}$) 
improvement in continuum sensitivity from the increased recording capability 
would lead to an average decrease in target-quasar separation by a factor of 
$>6$ (for standard $\log{N}/\log{S}$ statistics).  This should allow  proper 
motion accuracies of better than $\pm1$~\uas~year$^{-1}$ with less than 2 
years observations.

Combining the above mentioned advances in recording capabilities
with increased collecting area would allow measurement of the proper motions of 
weak radio sources in other galaxies.   AGN in many galaxies outside the 
Local Group have been detected, typically below 100~\uJy.  Noise levels
approaching $\sim1$~\uJy\ could be achieved by combining 32 Gbps recordings
and modest increases in collecting area from a ``pathfinder'' project that 
would prototype and test antenna ``patches'' (with 5\% of an SKA collecting area).  
Placing some of these antenna patches at VLBA sites, and at some new sites
between the EVLA and the VLBA, would greatly increase the sensitivity of 
the array, not only for continuum but also for spectral-line sources.  
The critical goal of measuring the proper motion of the Andromeda galaxy (M~31), 
which has a compact, but faint nuclear source of about 30~\uJy, could be
attempted with the recording rate upgrade (with about 30 tracks per epoch)
and easily accomplished (with a few tracks per epoch) with both the recording rate 
and collecting area upgrades.  

Some of these telescope advances discussed above could come from the phased 
implementation of the plans outlined in the ``North American Array'' initiative 
(J. Ulvestad, coordinator) submitted to the Decadal Survey.  In addition,
in order to map southern sources, we will need a high-sensitivity 
VLBA-like capability in the southern hemisphere.   This could be well met by 
a prototype (\eg 5\%) SKA, which is planned for the southern hemisphere.
Finally, we note that the construction of the full SKA would revolutionize 
all of these activities and lead to truly dramatic astrometric results well 
beyond the Virgo Cluster.

\section {Summary}

Over the next decades, a program of observational advances
could enable radio astrometry to address the following key questions:
\begin{itemize}
\item{} What is the distribution of dark matter in the Local Group?
\item{} What is the history and fate of Local Group galaxies?
\item{} What is the distribution of dark matter in nearby clusters of galaxies?
\end{itemize}

\begin{deluxetable}{ll}
\tablecolumns{2} \tablewidth{0pc} 
\tablecaption{Telescopes Advances and their Scientific Impact}
\tablehead {
  \colhead{Telescope Advance} & \colhead{Scientific Impact} 
            }
\startdata
High (32 Gbps) data recording rate and/or &
Background calibrators $\times6$ nearer to targets, \\
additional telescopes/collecting area
& enabling sub-\uas\ astrometry \\
\\
&
Proper motions of weak ($\sim10$~\uJy) AGNs \\
&
in Local Group and beyond \\
&
(\eg Andromeda, M81, Virgo Cluster) \\
\\
Improved Southern Hemisphere VLBI &
Proper motions and geometric distances \\
capability (\eg SKA) 
&of southern galaxies (\eg LMC, SMC)\\
\enddata
\label{table:telescope_impact}
\end{deluxetable}
\end{document}